\documentclass[12pt,journal,final,onecolumn]{IEEEtran}

\makeatletter
\let\IEEEproof\proof
\let\IEEEendproof\endproof
\let\proof\@undefined
\let\endproof\@undefined
\makeatother

\usepackage[dvips]{graphicx}
\usepackage{epsfig}
\usepackage{amsmath}
\usepackage{amssymb}
\usepackage{amsthm}
\usepackage{amsfonts}
\usepackage{bm}
\usepackage{color}
\usepackage[mathscr]{eucal}

\let\proof\IEEEproof
\let\endproof\IEEEendproof

\newcommand{\mc}[1]{\mathcal{#1}}

\newcommand{\mbb}[1]{\mathbb{#1}}

\newcommand{\defeq}{\triangleq}
\newcommand{\Pp}{\mathbb{P}}

\newcommand{\ind}{1\hspace{-2.5mm}{1}}

\newcommand{\Overline}[1]{\overline{\overline{#1}}}

\newcommand{\argmin}{\mathop{\rm arg~min}\limits}
\newcommand{\conv}{\mathrm{conv}}

\newcommand{\R}{\mathbb{R}}

\newcommand{\beq}{\begin{eqnarray}}
\newcommand{\eeq}{\end{eqnarray}}
\newcommand{\beqn}{\begin{eqnarray*}}
\newcommand{\eeqn}{\end{eqnarray*}}
\newcommand{\bequ}{\begin{equation}}
\newcommand{\eequ}{\end{equation}}
\newcommand{\bequn}{\begin{equation*}}
\newcommand{\eequn}{\end{equation*}}

\newcommand{\cP}{\mc{P}}
\newcommand{\cQ}{\mc{Q}}
\newcommand{\cA}{\mc{A}}
\newcommand{\cB}{\mc{B}}
\newcommand{\cM}{\mc{M}}

\newcommand{\cG}{\mc{G}}
\newcommand{\cX}{\mc{X}}

\newcommand{\hA}{\widehat{A}}

\newcommand{\hQ}{\widehat{Q}}

\newcommand{\Sg}{\Sigma}
\newcommand{\sg}{\sigma}

\newcommand{\beps}{\varepsilon}

\newtheorem{lemma}{Lemma}

\newtheorem{thm}[lemma]{Theorem}
\newtheorem{theorem}[lemma]{Theorem}

\theoremstyle{definition}
\newtheorem{egdummy}{Example}

\theoremstyle{remark}
\newtheorem{remark}{Remark}

\begin{document}

\bibliographystyle{unsrt}

\title{Adaptive Alternating Minimization Algorithms} 

\author{Urs Niesen, Devavrat Shah, Gregory Wornell%
\thanks{This work was supported in part by NSF under Grant
No.~CCF-0515109, and by HP through the MIT/HP Alliance.}
\thanks{The material in this paper was presented in part at the
IEEE International Symposium on Information Theory, Nice, June 2007.}
\thanks{The authors are with the Massachusetts Institute of Technology,
Department of Electrical Engineering and Computer Science, Cambridge,
MA. Email: \{uniesen,devavrat,gww\}@mit.edu}
}

\maketitle

\begin{abstract}
   The classical alternating minimization (or projection) algorithm has
   been successful in the context of solving optimization problems over
   two variables. The iterative nature and simplicity of the algorithm
   has led to its application in many areas such as signal processing,
   information theory, control, and finance. 
   
   A general set of sufficient conditions for the convergence and
   correctness of the algorithm are known when the underlying problem
   parameters are fixed.  In many practical situations, however, the
   underlying problem parameters are changing over time, and the use of
   an adaptive algorithm is more appropriate.  In this paper, we study
   such an adaptive version of the alternating minimization algorithm.
   More precisely, we consider the impact of having a slowly
   time-varying domain over which the minimization takes place.  As a
   main result of this paper, we provide a general set of sufficient
   conditions for the convergence and correctness of the adaptive
   algorithm. Perhaps somewhat surprisingly, these conditions seem to be
   the minimal ones one would expect in such an adaptive setting. We
   present applications of our results to adaptive decomposition of
   mixtures, adaptive log-optimal portfolio selection, and adaptive
   filter design. 
\end{abstract}

\section{Introduction}

\subsection{Background}

Solving an optimization problem over two variables in a product space is
central to many applications in areas such as signal processing,
information theory, statistics, control, and finance. The alternating
minimization or projection algorithm has been extensively used in such
applications due to its iterative nature and simplicity.

The alternating minimization algorithm attempts to solve a 
minimization problem of the following form: given $\cP$, $\cQ$
and a function $D: \cP \times \cQ \to \R$, 
minimize $D$ over $\cP \times \cQ$. That is, find  
\begin{equation*}
    \label{eq:min_orig}
    \min_{(P,Q) \in \cP\times \cQ}  D(P,Q).
\end{equation*}
Often minimizing over both variables simultaneously is not
straightforward. However, minimizing with respect to one variable while
keeping the other one fixed is often easy and sometimes possible
analytically. In such a situation, the alternating minimization
algorithm described next is well suited: start with an arbitrary initial
point $Q_0\in\mc{Q}$; for $n \geq 1$, iteratively compute
\begin{equation}
    \label{eq:proj_orig}
    \begin{aligned}
        P_n & \in \argmin_{P\in\mc{P}} D(P,Q_{n-1}), \\
        Q_n & \in \argmin_{Q\in\mc{Q}} D(P_n,Q).
    \end{aligned}
\end{equation}
In other words, instead of solving the original minimization problem
over two variables, the alternating minimization algorithm solves a
sequence of minimization problems over only one variable.  If the
algorithm converges, the converged value is returned as the solution to
the original problem. Conditions for the convergence and correctness
of such an algorithm, that is, conditions under which
\begin{equation}
    \label{eq:algconv}
    \lim_{n\to\infty}D(P_n,Q_n) 
    = \min_{(P,Q) \in \cP\times \cQ}  D(P,Q),
\end{equation}
have been of interest since the early 1950s. A general set of
conditions, stated in the paper by Csisz{\'a}r  and
Tusn{\'a}dy~\cite[Theorem 2]{ct}, is summarized in the next
theorem.\footnote{The conditions in \cite{ct} are actually slightly more
general than the ones shown here and allow for functions $D$ that take
the value $+\infty$, i.e.,
$D:\mbb{R}\times\mbb{R}\to\mbb{R}\cup\{+\infty\}$.}
\begin{theorem}\label{thm:zero}
    Let $\cP$ and $\cQ$ be any two sets, and  let $D: \cP \times \cQ \to
    \mbb{R}$ such that for all $\widetilde{P}\in\cP$, $\widetilde{Q}\in\cQ$
    \begin{align*}
        \argmin_{P\in\mc{P}} D(P,\widetilde{Q}) & \neq \emptyset, \\
        \argmin_{Q\in\mc{Q}} D(\widetilde{P},Q) & \neq \emptyset.
    \end{align*}
    Then the alternating minimization algorithm converges, i.e.,
    (\ref{eq:algconv}) holds, if there exists a nonnegative function
    $\delta:\cP \times \cP \to\mbb{R}_+$ such that the
    following two properties hold:
    \begin{itemize}
        \item[(a)] Three point property ($P, \widetilde{P}, \widetilde{Q}$): For all
            $P\in\mc{P}$, $\widetilde{Q}\in \mc{Q}$, $\widetilde{P}\in\argmin_{P\in\mc{P}}D(P,\widetilde{Q}),$
            \begin{equation*}
                \delta(P,\widetilde{P})+D(\widetilde{P},\widetilde{Q})   \leq D(P,\widetilde{Q}).
            \end{equation*}
        \item[(b)] Four point property ($P, Q, \widetilde{P}, \widetilde{Q}$): 
            For all $P,\widetilde{P}\in\mc{P}$, $Q\in\mc{Q}$, $\widetilde{Q}\in
            \argmin_{Q\in\mc{Q}} D(\widetilde{P},Q)$,
            \begin{equation*}
                D(P,\widetilde{Q}) \leq D(P,Q)+\delta(P,\widetilde{P}).
            \end{equation*}
    \end{itemize}
\end{theorem}

\subsection{Our Contribution}

In this paper, we consider an adaptive version of the above minimization
problem. As before, suppose we wish to find 
\begin{equation*}
    \min_{(P,Q) \in \cP\times \cQ} D(P,Q)
\end{equation*}
by means of an alternating minimization algorithm. However, on the
$n$\/th iteration of the algorithm, we are provided with sets $\cP_n, \cQ_n$ which are
{\em time-varying} versions of the sets $\cP$ and $\cQ$, respectively.  That
is, we are given a sequence of optimization problems
\begin{equation}\label{eq:min_seq}
    \Big\{\min_{(P,Q)\in\mc{P}_n\times\mc{Q}_n}D(P,Q)\Big\}_{n\geq 0} .
\end{equation}
Such situations arise naturally in many applications.  For example, in
adaptive signal processing problems, the changing parameters could be
caused by a slowly time-varying system, with the index $n$ representing
time. An obvious approach is to solve each of the problems
in~\eqref{eq:min_seq} independently (one at each time instance $n$).
However, since the system varies only slowly with time, such an approach
is likely to result in a lot of redundant computation. Indeed, it is
likely that a solution to the problem at time instance $n-1$ will be
very close to  the one at time instance $n$. A different approach is to
use an \emph{adaptive} algorithm instead. Such an adaptive algorithm
should be computationally efficient: given the tentative solution at
time  $n-1$, the tentative solution at time $n$ should be easy to
compute. Moreover,  if the time-varying system eventually reaches steady
state, the algorithm should converge to the optimal steady state
solution. In other words, instead of insisting that the adaptive
algorithm solves~\eqref{eq:min_seq} for every $n$, we only impose that
it does so as $n\to\infty$.

Given these requirement, a natural candidate for such an algorithm
is the following adaptation of the alternating minimization algorithm: 
start with an arbitrary initial $Q_0\in\mc{Q}_0$;
for $n \geq 1$ compute (cf.~\eqref{eq:proj_orig})
\begin{align*}
    \label{eq:adaptive}
    P_n & \in \argmin_{P\in\mc{P}_n} D(P,Q_{n-1}), \\
    Q_n & \in \argmin_{Q\in\mc{Q}_n} D(P_n,Q).
\end{align*}
Suppose that the sequences of sets $\{\mc{P}_n\}_{n\geq 0}$ and
$\{\mc{Q}_n\}_{n\geq 0}$ converge (in a sense to be made precise later)
to sets $\mc{P}$ and $\mc{Q}$, respectively. We are interested in
conditions under which 
\begin{equation*}
    \lim_{n\to\infty}D(P_n,Q_n) = \min_{(P,Q)\in\mc{P}\times\mc{Q}}D(P,Q).
\end{equation*}
As a main result of this paper, we provide
a general set of sufficient conditions under which this
adaptive algorithm converges. These conditions are essentially
the same as those of \cite{ct} summarized in Theorem \ref{thm:zero}.
The precise results are stated in Theorem \ref{thm:main1}.

\subsection{Organization}

The remainder of this paper is organized as follows. In
Section~\ref{sec:setup}, we introduce notation, and some preliminary
results. Section \ref{sec:general} provides a convergence result for a
fairly general class of adaptive alternating minimization algorithms. We
specialize this result to adaptive minimization of divergences in
Section~\ref{sec:divergence},  and to adaptive minimization procedures
in Hilbert spaces (with respect to inner product induced norm) in
Section~\ref{sec:hilbert}.  This work was motivated by several
applications in which the need for an adaptive alternating minimization
algorithm arises.  We present an application in the divergence
minimization setting from statistics and finance in
Section~\ref{sec:divergence}, and an application in the Hilbert space
setting from adaptive signal processing in Section \ref{sec:hilbert}.
Section~\ref{sec:conclusion} contains concluding remarks.

\section{Notations and Technical Preliminaries}
\label{sec:setup}

In this section, we setup notations and present technical preliminaries
needed in the remainder of the paper.  Let $(\mc{M}, d)$ be a compact
metric space. Given two sets $\cA, \cB \subset \cM$, define the
\emph{Hausdorff distance} between them as 
\begin{equation*}
    d_H(\cA,\cB)
    \defeq \max\left\{\sup_{A\in\mc{A}}\inf_{B\in\mc{B}}d(A,B),
    \sup_{B\in\mc{B}}\inf_{A\in\mc{A}}d(A,B)\right\}.
\end{equation*}
It can be shown the $d_H$ is a metric, and in particular satisfies the
triangle inequality.

Consider a continuous function $D: \cM \times \cM \to \R$.  For compact sets
$\cA, \cB \subset \cM$, define the set
\begin{equation*}
    \cG(\cA, \cB) \defeq \argmin_{(A, B) \in \cA \times \cB} D(A,B).
\end{equation*}
With slight abuse of notation, let 
\begin{equation*}
    D(\mc{A},\mc{B})\defeq\min_{(A,B)\in\mc{A}\times\mc{B}} D(A,B).
\end{equation*}
Due to compactness of the sets $\cA, \cB$ and continuity
of $D$, we have $\cG(\cA,\cB) \neq \emptyset$, and hence
$D(\mc{A},\mc{B})$ is well-defined.

\subsection{Some Lemmas}

Here we state a few auxiliary lemmas used in the following. 

\begin{lemma}[{\cite[Lemma 1]{ct}}]
    \label{lem:basic}
    Let $\{a_n\}_{n\geq 0}, \{b_n\}_{n\geq 0}$
    be sequences of real numbers, satisfying 
    \begin{equation*}
        a_n + b_n \leq b_{n-1} + c
    \end{equation*}
    for all $n\geq 1$ and some $c\in\mbb{R}$.  If 
    $\limsup_{n\to\infty} b_n > - \infty$ then
    \begin{equation*}
        \liminf_{n\to\infty} a_n \leq c.
    \end{equation*}
    If, in addition\footnote{We use $(x)^+\defeq\max\{0,x\}$.},
    \begin{equation*}
        \sum_{n=0}^{\infty} (c-a_n)^+ < \infty
    \end{equation*}
    then 
    \begin{equation*}
        \lim_{n\to\infty}a_n = c.
    \end{equation*}
\end{lemma}

\begin{lemma}
    \label{lem:prelim}
    Let $\{\mc{A}_n\}_{n\geq 0}$ be a sequence of subsets of $\cM$. Let
    $\cA$ be a closed subset of $\cM$ such that  $ \cA_n
    \stackrel{d_H}{\to} \cA.$ Consider any sequence $\{A_n\}_{n\geq 0}$
    such that $A_n \in \cA_n$ for all $n\geq 0$, and such that $A_n
    \stackrel{d}{\to} A\in\cM.$ Then $A \in \cA$. 
\end{lemma}
\begin{proof}
    Since $A_n \in \cA_n$ and $\cA_n \stackrel{d_H}{\to} \cA$,
    the definition of Hausdorff distance implies that there exists a sequence
    $\{\hA_n\}_{n\geq 0}$ such that $\hA_n \in \cA$ for all $n$ and 
    $d(\hA_n, A_n) \to 0$ as $n\to\infty$. Therefore
    \begin{equation*}
        d(\hA_n, A) \leq d(\hA_n, A_n) + d(A_n, A) \to 0
    \end{equation*}
    as $n\to\infty$. Since the sequence $\{\hA_n\}_{n\geq 0}$ is
    entirely in $\cA$, this implies that $A$ is a limit point of $\cA$.
    As $\cA$ is closed, we therefore have $A \in \cA$. 
\end{proof}

Let $(\cX,d)$ be a metric space and $f:\cX\to\R$. Define the \emph{modulus of
continuity} $\omega_f:\mbb{R}_+\to\mbb{R}_+$ of $f$ as
\begin{equation*}
    \omega_f(t) \defeq \sup_{\substack{x,x'\in\cX:\\d(x,x')\leq t}}
    \vert f(x)-f(x')\vert.
\end{equation*}

\begin{remark}
    \label{rem:uc}
    Note that if $f$ is uniformly continuous then $w_f(t)\to 0$ as $t\to
    0$. In particular, if $(\cX,d)$ is compact and $f$ is continuous then
    $f$ is uniformly continuous, and hence $\lim_{t\to 0}w_f(t)= 0$.
\end{remark}

\section{Adaptive Alternating Minimization Algorithms}
\label{sec:general}

Here we present the precise problem formulation. We then present an adaptive
algorithm and sufficient conditions for its convergence and correctness.

\subsection{Problem Statement}

Consider a compact metric space $(\cM, d)$, compact sets $\cP, \cQ
\subset \cM$, and a continuous function $D: \cM \times \cM \to \R$. We
want to find $D(\mc{P},\mc{Q})$. However, we are not given the sets
$\cP, \cQ$ directly. Instead, we are given a sequence of compact sets $\{(\cP_n,
\cQ_n)\}_{n\geq 0}$: $\cP_n, \cQ_n \subset\cM$ are revealed at time $n$
such that as $n\to \infty$, $\cP_n \stackrel{d_H}{\to} \cP$ and $\cQ_n
\stackrel{d_H}{\to} \cQ$. Given an arbitrary initial  
$(P_0, Q_0) \in \cP_0 \times \cQ_0$, the goal is to find a sequence of
points $(P_n, Q_n) \in \cP_n \times \cQ_n$ such that 
\begin{equation*}
    \lim_{n\to\infty}D(P_n,Q_n) 
    = D(\mc{P},\mc{Q}).
\end{equation*}

\subsection{Algorithm}

The problem formulation described in the last section suggests the following adaptive
version of the alternating minimization algorithm.
Initially, we have $(P_0, Q_0) \in\cP_0 \times \mc{Q}_0$. Recursively
for $n \geq 1$, pick any
\begin{align*}
    P_n & \in\argmin_{P\in\mc{P}_n} D(P,Q_{n-1}), \\
    Q_n & \in\argmin_{Q\in\mc{Q}_n}D(P_n,Q).
\end{align*}
We call this the Adaptive Alternating Minimization (AAM) algorithm in
the sequel.  Note that if $\cP_n = \cP$ and $\cQ_n = \cQ$ for all
$n$, then the above algorithm specializes to the classical alternating
minimization algorithm.  

\subsection{Sufficient Conditions for Convergence}

In this section, we present a set of sufficient conditions under which
the AAM algorithm  converges to $D(\mc{P},\mc{Q})$.  As we shall
see, we need ``three point'' and ``four point'' properties (generalizing
those in \cite{ct}) also in the adaptive setup.  To this end, assume
there exists a function\footnote{Note that unlike the condition in
\cite{ct}, we do not require $\delta$ to be nonnegative here.}
$\delta: \cM \times \cM \to \R$ such
that the following conditions are satisfied.
\begin{itemize}
    \item[(C1)] {\em Three point property} ($P, \widetilde{P}, Q$): for all $n \geq 1$,
        $P\in\mc{P}_n$, $Q\in\mc{Q}_{n-1}$,
        $\widetilde{P}\in\argmin_{P\in\mc{P}_n}D(P,Q)$,
        \begin{equation*}
            \delta(P,\widetilde{P})+D(\widetilde{P},Q)   
            \leq D(P,Q).
        \end{equation*}
    \item[(C2)] {\em Four point property} ($P, Q, \widetilde{P}, \widetilde{Q}$): for all $n \geq 1$,
        $P,\widetilde{P}\in\mc{P}_n$, $Q\in\mc{Q}_n$,
        $\widetilde{Q}\in\argmin_{Q\in\mc{Q}_n}D(\widetilde{P},Q)$,
        \begin{equation*}
            D(P,\widetilde{Q}) 
            \leq D(P,Q)+\delta(P,\widetilde{P}).
        \end{equation*}
\end{itemize}

Our main result is as follows.
\begin{thm}
    \label{thm:main1}
    Let $\{(\mc{P}_n, \mc{Q}_n)\}_{n\geq 0},\mc{P},\mc{Q}$ 
    be compact  subsets of the compact metric space $(\mc{M},d)$ such that
    \begin{equation*}
        \mc{P}_n\stackrel{d_H}{\to}\mc{P},\quad\quad
        \mc{Q}_n\stackrel{d_H}{\to}\mc{Q},
    \end{equation*}
    and let $D:\mc{M}\times\mc{M}\to\mbb{R}$ be a continuous function.
    Let conditions C1 and C2 hold.  Then, under the
    AAM algorithm,
    \begin{equation*}
        \liminf_{n\to\infty} D(P_n, Q_n) = D(\cP,\cQ),
    \end{equation*}
    and all limit points of subsequences of
    $\{(P_n, Q_n)\}_{n\geq 0}$ achieving this lim inf
    belong to $\mc{G}(\cP,\cQ)$. If, in addition,
    \begin{equation*}
        \sum_{n=0}^{\infty} \omega(2\varepsilon_n) < \infty,
    \end{equation*}
    where $\varepsilon_n\defeq
    d_H(\mc{P}_n,\mc{P})+d_H(\mc{Q}_n,\mc{Q})$, and
    $\omega\defeq\omega_D$ is the modulus of continuity of $D$, then
    \begin{equation*}
        \lim_{n\to\infty}D(P_n,Q_n) = D(\cP,\cQ),
    \end{equation*}
    and all limit points of $\{(P_n,Q_n)\}_{n\geq 0}$ belong to
    $\mc{G}(\mc{P},\mc{Q})$.
\end{thm}

\vspace{.1in}
\begin{remark} 
    Compared to the conditions of \cite[Theorem 2]{ct} summarized in Theorem 
    \ref{thm:zero}, the main additional
    requirement here is in essence uniform continuity of the function
    $D$ (which is implied by compactness of $\cM$ and continuity of
    $D$), and summability of the $\omega(2\varepsilon_n)$. This is the
    least one would expect in this adaptive setup to obtain a conclusion
    as in Theorem \ref{thm:main1}.
\end{remark}

\subsection{Proof of Theorem \ref{thm:main1}}

We start with some preliminaries. Given that $(\cM,d)$ is compact, the 
product space $(\cM \times \cM, d_2)$ with 
\begin{equation*}
    d_2((A, B), (A', B')) \defeq d(A, A') + d(B, B')
\end{equation*}
for all $(A, B), (A', B') \in \cM \times \cM$, is compact. Let
$\omega:\mbb{R}_+\to\mbb{R}_+$ be the modulus of continuity of $D$ with
respect to the metric space $(\cM\times\cM, d_2)$. By definition of
$\omega$, for any $\beps > 0$ and $(A, B), (A', B') \in \cM \times \cM$
such that 
\begin{equation*}
    d_2((A, B), (A', B')) \leq \beps,
\end{equation*}
we have
\begin{equation*}
    |D(A, B) - D(A',B')| \leq \omega(\beps).
\end{equation*}
Moreover, continuity of $D$ and compactness of $\cM
\times \cM$ imply (see Remark~\ref{rem:uc}) that $\omega(\beps) \to 0$
as $\beps \to 0$.

Recall the definition of  
\begin{equation*}
    \beps_n \defeq d_H(\cP_n, \cP) + d_H(\cQ_n, \cQ). 
\end{equation*}
By the hypothesis of Theorem~\ref{thm:main1}, we have $\beps_n
\to 0$ as $n\to\infty$, and
\begin{equation*}
    d_H(\cP_n, \cP_{n-1}) + d_H(\cQ_n, \cQ_{n-1}) 
    \leq \beps_{n-1} + \beps_n \defeq \gamma_n,
\end{equation*}
with $\gamma_n
\to 0$ as $n\to\infty$. 

We now proceed to the proof of Theorem \ref{thm:main1}. Condition C1
implies that for all $n\geq 1$, $P \in \cP_n, Q \in \cQ_n$,
\begin{equation}
    \label{eq:f0a}
    \delta(P,P_n)+D(P_n,Q_{n-1})
    \leq D(P,Q_{n-1}).
\end{equation}
Condition C2 implies that for all $n\geq 1$, $P \in \cP_n, Q \in \cQ_n$,
\begin{equation}
    \label{eq:f0b}
    D(P,Q_n)
    \leq D(P,Q)+\delta(P,P_n).
\end{equation}
Adding \eqref{eq:f0a} and \eqref{eq:f0b}, we obtain that 
for all $n\geq 1$, $P \in \cP_n, Q \in \cQ_n$,
\begin{equation}
    \label{eq:f1}
    D(P_n, Q_{n-1}) + D(P, Q_{n}) 
    \leq D(P, Q_{n-1}) + D(P, Q).
\end{equation}
Given that $d_H(\cQ_{n-1}, \cQ_n) \leq \gamma_n$, there exists $\hQ_n
\in \cQ_n$ such $d(Q_{n-1}, \hQ_n) \leq \gamma_n$. It follows that
\begin{equation*}
    d_2((P_n,\hQ_n), (P_n, Q_{n-1})) \leq \gamma_n,
\end{equation*}
and hence
\begin{equation}
    \label{eq:f2}
    \big\vert D(P_n, \hQ_n) - D(P_n, Q_{n-1})\big\vert   
    \leq \omega(\gamma_n).
\end{equation}
From (\ref{eq:f2}) and the AAM algorithm, we have 
\begin{equation}
    \label{eq:f3}
    \begin{aligned}
        D(P_n, Q_n) & = \min_{Q \in \cQ_n} D(P_n, Q) & \\
        & \leq D(P_n, \hQ_n) & (\text{since $\hQ_n \in \cQ_n$}) \\
        & \leq D(P_n, Q_{n-1}) + \omega(\gamma_n).
    \end{aligned}
\end{equation}
Adding inequalities \eqref{eq:f1} and \eqref{eq:f3},
\begin{equation}
    \label{eq:f4}
    D(P_n, Q_{n}) + D(P, Q_{n}) 
    \leq D(P, Q_{n-1}) + D(P, Q) + \omega(\gamma_n),
\end{equation}
for all $P \in \cP_n, Q\in \cQ_n$. 

Since $\cP_n\stackrel{d_H}{\rightarrow}\cP$ and
$\cQ_n\stackrel{d_H}{\rightarrow}\cQ$, there exists a sequence $(P^*_n,
Q^*_n) \in \cP_n \times \cQ_n$ such that $(P^*_n, Q^*_n) \to (P^*, Q^*)
\in \cG(\cP, \cQ)$ and $d_2((P^*_n, Q^*_n), (P^*,Q^*)) \leq \beps_n$ for
all $n\geq 0$. Pick any such sequence $\{(P^*_n, Q^*_n)\}_{n\geq 0}$.
Replacing $(P,Q)$ in \eqref{eq:f4} by this $(P^*_n, Q^*_n)$, we obtain
\begin{equation}
    \label{eq:f5}
    D(P_n, Q_{n}) + D(P^*_n, Q_{n}) 
    \leq D(P^*_n, Q_{n-1}) + D(P^*_n, Q^*_n) + \omega(\gamma_n). 
\end{equation}
By choice of the $(P^*_n,Q^*_n)$,
\begin{equation}
    \label{eq:f6a}
    D(P^*_n, Q^*_n) 
    \leq D(P^*, Q^*) + \omega(\varepsilon_n).
\end{equation}
Moreover,  
\begin{align*}
    d(P^*_{n-1}, P^*_n) 
    & \leq d(P^*_{n-1}, P^*)+d(P^*, P^*_n) \\
    & \leq \varepsilon_{n-1}+\varepsilon_n \\
    & = \gamma_n,
\end{align*}
and therefore
\begin{equation}
    \label{eq:f6b}
    D(P^*_n, Q_{n-1}) 
    \leq D(P^*_{n-1}, Q_{n-1}) + \omega(\gamma_n).
\end{equation}
Combining inequalities \eqref{eq:f6a} and \eqref{eq:f6b} with \eqref{eq:f5}, we obtain
\begin{equation}
    \label{eq:f7}
    D(P_n, Q_{n}) + D(P^*_n, Q_{n}) 
    \leq D(P^*_{n-1}, Q_{n-1}) + D(P^*, Q^*) + 2\omega(\gamma_n)+\omega(\varepsilon_n).
\end{equation}
Define 
\begin{align*}
    a_n & \defeq D(P_n,Q_n)-2\omega(\gamma_n)-\omega(\varepsilon_n), \\
    b_n & \defeq D(P^*_n, Q_n), \\
    c & \defeq D(P^*, Q^*),
\end{align*}
and note that by \eqref{eq:f7}
\begin{equation*}
    a_n+b_n \leq b_{n-1}+c.
\end{equation*}
Since $D$ is a continuous function over the compact set $\cM \times
\cM$, it is also a bounded function.  Hence we have
$\limsup_{n\to\infty} \vert b_n \vert< \infty$.  Applying Lemma
\ref{lem:basic},
\begin{equation}
    \label{eq:f8}
    \liminf_{n\to\infty} D(P_n, Q_n) 
    \leq D(P^*, Q^*) 
    + \limsup_{n\to\infty}
    \big(2\omega(\gamma_n)+\omega(\varepsilon_n)\big).
\end{equation}
Since $\gamma_n \to 0$ and $\varepsilon_n\to 0$ imply
$2\omega(\gamma_n)+\omega(\varepsilon_n)\to 0$,
\eqref{eq:f8} yields 
\begin{equation}
    \label{eq:upper}
    \liminf_{n\to\infty} D(P_n, Q_n) \leq D(\cP, \cQ). 
\end{equation}
Now, let $\{n_k\}_{k\geq 0}$ be a subsequence such that
\begin{equation*}
    \liminf_{n\to\infty} D(P_n, Q_n)  
    = \lim_{k\to\infty} D(P_{n_k}, Q_{n_k}).
\end{equation*}
By compactness of $\mc{M}\times\mc{M}$, we can assume without loss of
generality that $P_{n_k}\stackrel{d}{\to}P$, $Q_{n_k}\stackrel{d}{\to}Q$ for some
$P,Q\in\mc{M}$. Since $\mc{P}$ and $\mc{Q}$ are compact, 
Lemma~\ref{lem:prelim} shows that $P\in\mc{P}$, $Q\in\mc{Q}$.
By continuity of $D$ this implies that
\begin{align*}
    \liminf_{n\to\infty} D(P_n, Q_n)
    & = \lim_{k\to\infty} D(P_{n_k}, Q_{n_k}) \\
    & = D(P,Q) \\
    & \geq D(\cP, \cQ).
\end{align*}
Together with~\eqref{eq:upper}, this shows that
\begin{equation*}
    \liminf_{n\to\infty} D(P_n, Q_n) 
    = D(\cP, \cQ),
\end{equation*}
and that all limit points of subsequences of $\{(P_n,Q_n)\}_{n\geq 0}$
achieving this lim inf belong to $\mc{G}(\mc{P},\mc{Q})$.
This completes the proof the first part of Theorem \ref{thm:main1}.

Suppose now that we have in addition
\begin{equation}
    \label{eq:f9}
    \sum_{n=0}^{\infty} \omega(2\varepsilon_n) < \infty.
\end{equation}
Since
\begin{align*}
    D(P_n,Q_n) 
    & \geq \min_{P\in\mc{P}_n,Q\in\mc{Q}_n} D(P,Q) \\
    & \geq \min_{P\in\mc{P},Q\in\mc{Q}} D(P,Q)-\omega(\varepsilon_n) \\
    & = D(P^*,Q^*)-\omega(\varepsilon_n),
\end{align*}
we have
\begin{align*}
    (c-a_n)^+ 
    & = \big( D(P^*,Q^*)-D(P_n,Q_n)+2\omega(\gamma_n)+\omega(\varepsilon_n) \big)^+ \\
    & \leq 2\big( \omega(\gamma_n)+\omega(\varepsilon_n) \big) \\
    & \leq 2\big(\omega(2\varepsilon_n)+\omega(2\varepsilon_{n-1})+\omega(\varepsilon_n) \big) \\
    & \leq 2\big(2\omega(2\varepsilon_n)+\omega(2\varepsilon_{n-1})\big).
\end{align*}
Thus by~\eqref{eq:f9}, 
\begin{equation*}
    \sum_{n=0}^{\infty}(c-a_n)^+ < \infty,
\end{equation*}
and applying again Lemma~\ref{lem:basic} yields
\begin{equation}
    \label{eq:don}
    \lim_{n\to\infty}D(P_n,Q_n) = D(P^*,Q^*).
\end{equation}
As every limit point of $\{(P_n,Q_n)\}_{n\geq 0}$ belongs to
$\mc{P}\times\mc{Q}$ by Lemma~\ref{lem:prelim}, \eqref{eq:don} and
continuity of $D$ imply that if~\eqref{eq:f9} holds then every limit
point of $\{(P_n,Q_n)\}_{n\geq 0}$ must also belong to
$\mc{G}(\mc{P},\mc{Q})$.  This concludes the proof of
Theorem~\ref{thm:main1}.

\section{Divergence Minimization}
\label{sec:divergence}

In this section, we specialize the algorithm from
Section~\ref{sec:general} to the case of alternating divergence
minimization. A large class of problems can be formulated as a
minimization of divergences.  For example, computation of channel
capacity and rate distortion function~\cite{ari, bla}, selection of 
log-optimal portfolios~\cite{cov}, and maximum likelihood estimation from
incomplete data~\cite{dem}. These problems were shown to be divergence
minimization problems in~\cite{ct}. For further applications of
alternating divergence minimization algorithms, see~\cite{sul}.  We
describe applications to the problem of adaptive mixture decomposition
and of adaptive log-optimal portfolio selection.

\subsection{Setting}
\label{sec:divsetup}

Given a finite set $\Sg$ and some constant $0 < b  < B$, let
$\cM = \cM(\Sg, b, B)$ be the set of  all measures $P$ on
$\Sigma$ such that 
\begin{equation}
    \label{eq:mbounds}
    \sum_{\sigma \in \Sg} P(\sg) \leq B,
    ~\mbox{and}~   P(\sigma) \geq b, ~\forall ~\sigma \in \Sg.
\end{equation}
Endow $\cM$ with the topology induced by the metric $d: \cM \times \cM
\to \R_+$ defined as
\begin{equation*}
    d(P,Q) 
    \defeq \max_{\sg \in\Sg}\vert P(\sg) - Q(\sg)\vert.
\end{equation*}
It is easy to check that the metric space $(\cM, d)$ is compact. 
The cost function $D$ of interest is divergence\footnote{All logarithms
are with respect to base $e$.}
\begin{equation*}
    D(P,Q) \defeq D(P\Vert Q) 
    \defeq \sum_{\sg \in \Sg} P(\sg) \log \frac{P(\sg)}{Q(\sg)}
\end{equation*}
for any $P, Q \in \cM$. Note that~\eqref{eq:mbounds} ensures that $D$ is
well defined (i.e., does not take the value $\infty$).  It is well-known
(and easy to check) that the function $D$ is
continuous and convex in both arguments. Finally, define the function
$\delta:\cM\times\cM\to\mbb{R}$
\begin{equation*}   
    \delta(P,\widetilde{P}) 
    \defeq D(P\Vert \widetilde{P})-\sum_{\sigma \in \Sg}\left(P(\sg)-\widetilde{P}(\sg)\right).
\end{equation*}

In \cite{ct}, it has been established that for convex $\mc{P}$ and
$\mc{Q}$ the pair of functions $D, \delta$ satisfy the ``three point''
and ``four point'' properties C1 and C2.  As stated above,
the space $\cM = \cM(\Sg, b, B)$ with metric $d$ is a compact metric
space, and the function $D$ is continuous. Hence Theorem~\ref{thm:main1} 
applies in this setting.

\subsection{Application: Decomposition of Mixtures and Log-Optimal
Portfolio Selection}

We consider an application of our adaptive divergence minimization
algorithm to the problem of decomposing a mixture. A special case of
this setting yields the problem of log-optimal portfolio selection.

We are given a sequence of i.i.d. random variables $\{Y_l\}_{l\geq 0}$,
each taking values in the finite set $\mc{Y}$. $Y_l$ is distributed
according to the mixture $\sum_{i=1}^{I}c_i\mu_i$, where the
$\{c_i\}_{i=1}^{I}$ sum to one, $c_i\geq c_0>0$ for all
$i\in\{1,\ldots, I\}$, and where $\{\mu_i\}_{i=1}^{I}$ are distributions
on $\mc{Y}$.  We assume that $\mu_i(y)\geq\mu_0>0$ for all $y\in\mc{Y},
i\in\{1,\ldots,I\}$.  The goal is to compute an estimate of
$\{c_i\}_{i=1}^I$ from $\{Y_l\}_{l=1}^{n}$ and knowing
$\{\mu_i\}_{i=1}^I$.

Let $\overline{P}_n:\mc{Y}\to[0,1]$,
\begin{equation*}
    \overline{P}_n(y)
    \defeq \frac{1}{n}\sum_{\ell=1}^n\ind_{\{Y_\ell=y\}},
\end{equation*}
be the empirical distribution of $\{Y_l\}_{l=1}^n$.
The maximum likelihood estimator of $\{c_i\}_{i=1}^I$ is given by (see,
e.g., ~\cite[Lemma 3.1]{cs})
\begin{equation}
    \label{eq:ml}
    \argmin_{\{\tilde{c}_i\}} 
    D\Big(\overline{P}_n\big\Vert{\textstyle\sum_{i=1}^I} \tilde{c}_i \mu_i\Big),
\end{equation}
Following~\cite[Example 5.1]{cs}, we define
\begin{equation}
    \label{eq:pndef}
    \begin{aligned}
        \Sigma & \defeq \{1,\ldots,I\}\times\mc{Y}, \\
        \mc{Q}_n & = \mc{Q} \defeq \{Q:Q(i,y)=\tilde{c}_i\mu_i(y), 
        \textrm{ for some $\{\tilde{c}_i\}$ with ${\textstyle\sum_{i}}\tilde{c}_i = 1,
        \tilde{c}_i\geq c_0\forall i$} \}, \\
        \mc{P}_n & \defeq \{P:{\textstyle\sum_{i=1}^I} P(i,y) = \overline{P}_n(y),
        P(i,y)\geq 0\forall i,y \}.
    \end{aligned}
\end{equation}
Note that $\mc{P}_n$ and $\mc{Q}$ are convex and compact. 
From~\cite[Lemma 5.1]{cs}, we have 
\begin{equation*}
    \min_{\{\tilde{c}_i\}}
    D\Big(\overline{P}_n\big\Vert{\textstyle\sum_{i=1}^I} \tilde{c}_i \mu_i\Big)
    = \min_{P\in\mc{P}_n}\min_{Q\in\mc{Q}}
    D(P\Vert Q),
\end{equation*}
and the minimizer of the left hand side (and hence \eqref{eq:ml}) is
recovered from the corresponding marginal of the optimal $Q$ on the
right hand side.

We now show how the projections on the sets $\mc{P}_n$ and $\mc{Q}$ can
be computed.  Fix a $P$, assuming without loss of generality that
\begin{equation*}
    \sum_{y\in\mc{Y}}P(1,y)
    \geq \sum_{y\in\mc{Y}}P(2,y)
    \geq \ldots
    \geq \sum_{y\in\mc{Y}}P(I,y).
\end{equation*}
We want to minimize $D(P\Vert Q)$ over all $Q\in\mc{Q}$, or,
equivalently, over all valid $\{\tilde{c_i}\}$.  The $\{\tilde{c_i}\}$
minimizing $D(P\Vert Q)$ can be shown to be of the form
$\tilde{c}_i>c_0$ for all $i\leq J^*$ and $\tilde{c}_i=c_0$ for all
$i>J^*$. More precisely, define
\begin{equation*}
    \eta(J) 
    \defeq \frac{1}{1-(I-J)c_0} \sum_{i=1}^J \sum_{y\in\mc{Y}}P(i,y),
\end{equation*}
and choose $J^*\in\{1,\ldots,I\}$ such that
\begin{align*}
    \frac{1}{\eta(J^*)}\sum_{y\in\mc{Y}}P(i,y) & > c_0 & \text{for $1\leq i\leq J^*$,} \\
    \frac{1}{\eta(J^*)}\sum_{y\in\mc{Y}}P(i,y) & \leq c_0 & \text{for $J^*< i\leq I$.} 
\end{align*}
Then the optimal $\{\tilde{c}_i\}$ are given by
\begin{align*}
    \tilde{c}_i & = \frac{1}{\eta(J^*)}\sum_{y\in\mc{Y}}P(i,y) & \text{for $1\leq i\leq J^*$,} \\
    \tilde{c}_i & = c_0 & \text{for $J^*< i\leq I$.} 
\end{align*}
For fixed $Q(i,y)=\tilde{c}_i\mu_i(y)$, the minimizing $P$ is
\begin{equation}
    \label{eq:minp}
    P(i,y) = \frac{\tilde{c}_i\mu_i(y)}{\sum_j\tilde{c}_j\mu_j(y)}
    \overline{P}_n(y).
\end{equation}

We now check that~\eqref{eq:mbounds} is satisfied for some values of $b$
and $B$. As $\mc{P}_n$ and $\mc{Q}$ are sets of distributions, we can
choose $B=1$.  For all $Q\in\mc{Q}$, $i\in\{1,\ldots,I\}$, $y\in\mc{Y}$,
we have $Q(i,y)\geq\mu_0 c_0>0$. However, for $P\in\mc{P}_n$, we have in
general only $P(i,y)\geq 0$. In order to apply the results from
Section~\ref{sec:divsetup}, we need to show that we can, without loss of
optimality, restrict the sets $\mc{P}_n$ to contain only distributions
$P$ that are bounded below by some $p_0>0$.  In other words, we need to
show that the projections on $\mc{P}_n$ are bounded below by $p_0$.

Assume for the moment that the empirical distribution $\overline{P}_n$ is
close to the true one in the sense that
\begin{equation*}
    \Big\vert\overline{P}_n(y)-\sum_i c_i\mu_i(y)\Big\vert 
    \leq \frac{\mu_0}{2}
\end{equation*}
for all $y\in\mc{Y}$.  As $\sum_i c_i\mu_i(y)\geq\mu_0$ this implies
$\overline{P}_n(y)\geq\frac{\mu_0}{2}$ for all $y$.
From~\eqref{eq:minp}, this implies that the projection $P$ in $\mc{P}_n$
of any point in $\mc{Q}$ satisfies $P(i,y)\geq\frac{1}{2}
c_0\mu_0^2\defeq p_0$ for all $i\in\{1,\ldots,I\}, y\in\mc{Y}$.  Hence
in this case $\mc{M}(\Sigma,b,B)$ satisfies~\eqref{eq:mbounds} with
$b=\frac{1}{2} c_0\mu_0^2$ and $B=1$.

It remains to argue that $\overline{P}_n$ is close to $\sum_i
c_i\mu_i(y)$. Suppose instead of constructing the set $\mc{P}_n$
(see~\eqref{eq:pndef}) with
respect to $\overline{P}_n$, we construct it with respect to the
distribution $\Overline{P}_n$ defined as
\begin{equation*}
    \Overline{P}_n(y) \defeq 
    \frac{\mu_0}{2}+\lambda\Big(\overline{P}_n(y)-\frac{\mu_0}{2}\Big)^+,
\end{equation*}
where $\lambda$ is chosen such that $\sum_{y}\Overline{P}_n(y)=1$.
$\Overline{P}_n$ is bounded below by $\frac{\mu_0}{2}$ by construction.
Moreover, by the strong law of large numbers,
\begin{equation*}
    \Pp(\overline{P}_n\neq\Overline{P}_n \textrm{ i.o.}) = 0.
\end{equation*}
Hence we have $\mc{P}_n\stackrel{d_H}{\to}\mc{P}$ almost surely, where
$\mc{P}$ is constructed as in~\eqref{eq:pndef} with respect to the 
true distribution $\sum_i c_i\mu_i$.

Applying now the results from Section~\ref{sec:divsetup} and
Theorem~\ref{thm:main1} yields that under the AAM algorithm
\begin{equation*}
    \liminf_{n\to\infty}D(P_n,Q_n) 
    = D(\mc{P},\mc{Q})
\end{equation*}
almost surely, and that every limit point of $\{(P_n,Q_n)\}_{n\geq 0}$ 
achieving this lim inf is an element of $\mc{G}(\mc{P},\mc{Q})$.

Since by the law of the iterated logarithm, convergence of
$\overline{P}_n$ to $P$ is only $\Theta(\sqrt{\log\log n}/\sqrt{n})$ as
$n\to\infty$ almost surely, and since $\lim_{\varepsilon\to
0}\omega(\varepsilon)/\varepsilon=0$ only if $D$ is a
constant~\cite{dev}, we can in this scenario \emph{not} conclude from
Theorem~\ref{thm:main1} that
$\lim_{n\to\infty}D(P_n,Q_n)=D(\mc{P},\mc{Q})$.

As noted in~\cite{cs}, a special case of the decomposition of mixture
problem is that of maximizing the expected value of $\log\sum_i c_i
W_i$, where $\{W_i\}_{i=1}^I$ is distributed according to
$\overline{P}_n$. The standard alternating divergence minimization
algorithm is then the same as Cover's portfolio optimization
algorithm~\cite{cov}. Thus the AAM algorithm applied as before yields
also an adaptive version of this portfolio optimization algorithm.

\section{Projections in Hilbert Space}
\label{sec:hilbert}

In this section, we specialize the algorithm from
Section~\ref{sec:general} to the case of minimization in a
Hilbert space. A large class of problems can be formulated as
alternating projections in Hilbert spaces. For example, problems in
filter design, signal recovery, and spectral estimation. For an
extensive overview, see~\cite{com1}. In the context of Hilbert spaces,
the alternating minimization algorithm is often called POCS (Projection
Onto Convex Sets). 

\subsection{Setting}
\label{sec:hilbert_setup}

Let $\mc{M}$ be a compact subset of a Hilbert space with the usual norm
$d(A,B)^2 \defeq \langle A-B,A-B\rangle$. Then $(\cM, d)$ is a compact
metric space.  The cost function $D$ of interest is
\begin{equation*}
    D(A,B) \defeq d(A,B)^2.
\end{equation*}
The function $D$ is continuous and convex. Define the
function $\delta$ (as part of conditions C1 and C2), as 
\begin{equation*}
    \delta(A,\widetilde{A})  \defeq d(A,\widetilde{A})^2.
\end{equation*}

In \cite{ct}, it is established that for convex $\mc{P}$ and 
$\mc{Q}$ the pair of functions $D,
\delta$ satisfies the ``three point'' and ``four point'' properties 
C1 and C2. Hence Theorem~\ref{thm:main1} applies in this
setting.

\subsection{Application: Set Theoretic Signal Processing and Adaptive Filter Design}

In this section, we consider a problem in the Hilbert space setting as
defined in Section~\ref{sec:hilbert_setup}. Let $\{\mc{S}_i\}_{i=1}^I$ be a
collection of convex compact subsets of the Hilbert space $\mbb{R}^k$
with the usual inner product, and let $\{c_i\}_{i=1}^I$ be positive
weights summing to one. In set-theoretic signal processing, the
objective is to find a point $A$ minimizing
\begin{equation}
    \label{eq:set}
    \sum_{i=1}^I c_i d(A,\mc{S}_i),
\end{equation}
where $d(A,\mc{S}_i)\defeq\min_{S\in\mc{S}_i}d(A,S)$. Many problems in
signal processing can be formulated in this way. Applications can be
found for example in control, filter design, and estimation. For
an overview and extensive list of references, see~\cite{com1}. As an
example, in a filter design problem, the $\mc{S}_i$ could be constraints
on the impulse and frequency responses of a filter~\cite{cet, nob}.

Following~\cite{com2}, this problem can be formulated in our framework by
defining the Hilbert space $\mc{H}=\mbb{R}^{Ik}$ with inner product
\begin{equation*}
    \langle A,B\rangle 
    \defeq \sum_{i=1}^I c_i \langle A_i,B_i\rangle,
\end{equation*}
where $A_i,B_i\in\mbb{R}^k$ for $i\in\{1,\ldots, I\}$ are the components
of $A$ and $B$. Let 
\begin{equation*}
    \mc{S}\defeq \conv\{\cup_{i=1}^I\mc{S}_i\}\subset\mbb{R}^k,
\end{equation*}
be the convex hull of the union of the constraint sets
$\{\mc{S}_i\}_{i=1}^I$, and let
\begin{equation*}
    \mc{M} \defeq \mc{S}^I\subset\mc{H}
\end{equation*}
be its $I$-fold product. Since each of the sets $\mc{S}_i$ is compact,
$\mc{M}$ is compact and by definition also convex.
We define the set $\mc{P}\subset\mc{M}$ as
\begin{equation*}
    \mc{P} \defeq \{(\widetilde{P},\ldots,\widetilde{P})\in\mc{H}: \widetilde{P}\in\mc{S}\}
\end{equation*}
and the set $\mc{Q}\subset\mc{M}$ as
\begin{equation}
    \label{eq:qdefs}
    \mc{Q} \defeq \mc{S}_1\times\cdots\times\mc{S}_I.
\end{equation}

We now show how the projections on the sets $\cP$ and $\cQ$ can be
computed. For a fixed $P=(\widetilde{P},\ldots,\widetilde{P})\in\mc{P}$,
the $Q\in\mc{Q}$ minimizing $D(P,Q)$ has the form 
\begin{equation*}
    \big(S_1(\widetilde{P}),\ldots,S_I(\widetilde{P})\big),
\end{equation*}
where $S_i(\widetilde{P})$ is the $Q_i\in\mc{S}_i$ minimizing
$\Vert\widetilde{P}-\widetilde{Q}_i\Vert^2$.
For a fixed $Q=(Q_1,\ldots,Q_I)\in\mc{Q}$ the $P\in\mc{P}$ minimizing 
$D(P,Q)$ is given by
\begin{equation*}
    \big({\textstyle\sum_{i=1}^I} c_i Q_i, \ldots, {\textstyle\sum_{i=1}^I} c_i Q_i\big).
\end{equation*}
Moreover, a solution to~\eqref{eq:set} can be found from the
standard alternating minimization algorithm for Hilbert spaces on
$\mc{P}$ and $\mc{Q}$.

To this point, we have assumed that the constraint sets
$\{\mc{S}_i\}_{i=1}^I$ are constant. The results from
Section~\ref{sec:general}, enable us to look at situations in which the
constraint sets $\{\mc{S}_{i,n}\}_{i=1}^I$ are time-varying.  Returning
to the filter design example mentioned above, we are now interested in
an adaptive filter. The need for such filters arises in many different
situations (see, e.g.,~\cite{hay}). 

The time-varying sets $\{\mc{S}_{i,n}\}_{i=1}^I$ give rise to sets
$\mc{Q}_n$, defined in analogy to~\eqref{eq:qdefs}.  We assume again
that $\mc{S}_{i,n}\stackrel{d_H}{\to}\mc{S}_i$ for all
$i\in\{1,\ldots,I\}$, and let $\mc{Q}$ be defined with respect to the
limiting $\{S_i\}_{i=1}^I$ as before. Applying the results from
Section~\ref{sec:hilbert_setup} and Theorem~\ref{thm:main1}, we obtain
convergence and correctness of the AAM algorithm.

\section{Conclusions}
\label{sec:conclusion}

We considered a fairly general adaptive alternating minimization
algorithm, and found sufficient conditions for its convergence and
correctness. This adaptive algorithm has applications in a variety of
settings. We discussed in detail how to apply it to three different
problems (from statistics, finance, and signal processing).

\section*{Acknowledgment}

The authors would like to thank the anonymous reviewer as well as the
associate editor Gerhard Kramer whose comments helped improving the
final version of this manuscript.

\bibliography{projections}

\end{document}